\documentclass[11pt, a4paper]{article}

\usepackage[usenames, dvipsnames]{color}
\usepackage[T1]{fontenc}
\usepackage[latin1]{inputenc}
\usepackage{lmodern}
\usepackage{amsmath}
\usepackage[pdftex]{graphicx}
\usepackage{lastpage}
\usepackage{amssymb} 
\usepackage{hyperref}

\usepackage{cite}

\usepackage{titling}
\usepackage{authblk}

\usepackage{subcaption}

\usepackage{yfonts}

\usepackage{esvect}

\usepackage{xspace}

\newcommand*{\fig}{Fig.\@\xspace}

\newcommand*{\Eq}{Eq.\@\xspace}

   \usepackage{hyphenat}   
   
\usepackage{slashed}
\usepackage{braket}
\usepackage{simplewick}
\usepackage{bbm}

\usepackage{mathtools}

\usepackage{booktabs}
\newcommand{\ra}[1]{\renewcommand{\arraystretch}{#1}}

\usepackage{xifthen}
\newcommand*\diff{\mathrm{d}} 
\newcommand*\ldiff[2][]{ \ifthenelse{\isempty{#1}}{ \diff #2}{\diff^#1#2} \,} 
\let\limitint\int 
\renewcommand{\int}{\limitint \!} 

\title{Numerical study of a cosmological relaxation model of the Higgs boson mass}
\author[1,2]{Marco Michel\thanks{michel@mpp.mpg.de}}
\affil[1]{Max-Planck-Institut für Physik, F\"ohringer Ring 6, 80805 M\"unchen, Germany}
\affil[2]{Arnold Sommerfeld Center, Ludwig-Maximilians-Universit\"at, \mbox{Theresienstraße 37, 80333 M\"unchen, Germany}}

\begin{document}
\allowdisplaybreaks

\maketitle
\begin{abstract}
In light of no discovered new physics at the LHC, ideas which tackle the Hierarchy Problem without novelties around the TeV scale
must be taken seriously.  Such is a cosmological relaxation model of the Higgs
mass, proposed in the pre-LHC era, which  does not rely on new physics below
the Planck scale. This scenario introduces a different notion of naturalness
according  to which the vacuum with a small expectation value of the Higgs field
corresponds to an infinitely enhanced entropy point of the vacuum landscape
that becomes an attractor of cosmological  inflationary evolution. In this framework
we study numerically the evolution of the Higgs VEV. We model the inflationary vacuum-to-vacuum transitions that are triggered by
nucleation of branes charged under three-form fields as a random walk.
In particular, we investigate the impact of the number of coupled
three-forms on the convergence rate of the Higgs VEV. We discover an
enhanced rate with increasing number of brane charges. Moreover, we show
that for late times the inclusion of more charges is equivalent to
additional brane nucleations.
\end{abstract}

\section{Introduction}
Rooted in the quadratic sensitivity of the Higgs mass term  to the UV
cutoff, the Hierarchy problem is a still unresolved puzzle in theoretical
physics.
(For discussion of the Hierarchy Problem and its connection to
naturalness, see \cite{D1908}). Naturally, since the standard picture raised the problem in the first place, a solution to it most likely has to contain new ingredients. Most of the proposed solutions such as supersymmetry, which has no quadratic divergences \cite{Nilles:1983ge},
technicolor \cite{Weinberg:1975gm},  extra dimensions \cite{ArkaniHamed:1998rs, Randall:1999ee} or large number of species \cite{Dvali:2007hz}
predict stabilizing new physics around a scale not much larger than the
weak scale. However, these predicted deviations from the standard model have not been observed at currently available energy scales.
Nonetheless, already in pre-LHC era an alternative idea of solving the
Hierarchy Problem via cosmological relaxation of the Higgs mass
has been suggested \cite{DV0304, D0410}.  This scenario does not rely on any low energy
physics as it can push the scale of the onset of new physics up to the Planck scale. It also introduces a notion of naturalness \cite{D1908}
that is fundamentally different from the standard one by 't Hooft \cite{tHooft:1979rat}.
Here, the vacuum with the small value of the Higgs mass has infinite
entropy
and represents an attractor of cosmological evolution.
The non-observation of any new physics at the LHC gives a serious
motivation for a detailed study of this scenario.
A list of recent references based on the same idea can be found in \cite{D1908}.

In this paper we first confirm the relaxation of the Higgs vacuum expectation value (VEV) to the
attractor
value by numerical simulations. Furthermore, we extend the analysis to
additional three-form fields and study their affect on the convergence rate.
This is
motivated by embedding the attractor scenario to various fundamental
theories that  contain multiple forms.

The paper is structured as follows. We first briefly review the cosmic
attractor model in section \ref{sec:review}. In chapter \ref{sec:rwModel} we map a simplified model to a
stochastic framework. The results and interpretation of the numerical
simulations are provided in section \ref{sec:main}. We conclude the discussion in
section \ref{sec:end}.

\section{Cosmic Higgs VEV Relaxation Mechanism}
\label{sec:review}
In this section we briefly review the Cosmic Attractor model introduced in \cite{DV0304} and further refined in \cite{D0410} and \cite{D1908}.
This model exhibits a high degeneracy of vacua around a certain hierarchically-small value of the Higgs VEV $\Phi_*$.
This point of \textit{enhanced entropy} is reached via brane nucleations during inflationary cosmological evolution. Since inflation last eternally \cite{Vilenkin:1983xq, Linde:1986fd}, the system has infinitely available time to converge to the attractor point. It is therefore \textit{natural} to find oneself in the vacuum with maximal entropy and corresponding Higgs VEV$\Phi_*$.

The crux of the mechanism is to couple the Higgs to a massless three-form field which is sourced by a 2-brane with charge $Q$ set by the Higgs field itself. The fundamental nature of the brane is not important, in particular as shown in \cite{DV0304} and \cite{D0410} it can be resolved in form of a domain wall of a heavy axion. Crossing the 2-brane (or axionic domain wall) leads to a jump in the field strength $F$ with the distance set by the brane charge:
\begin{equation}
\Delta F = Q(\Phi).
\label{eq:F}
\end{equation}
On the other hand, $F$ is back-controlling the vacuum expectation value of the Higgs field via
\begin{equation}
\Phi^2 = \frac{1}{\lambda} \left ( \frac{F^2}{M^2} - m^2\right),
\label{eq:Phi}
\end{equation}
where $\lambda$ is a coupling constant, $M$ is some cutoff and $m$ incorporates all other contributions to the effective Higgs mass. 
Using \Eq (\ref{eq:F}) and (\ref{eq:Phi}) the change of the Higgs VEV for small values of $Q$ when crossing a 2-brane is
\begin{equation}
\Phi \Delta \Phi = -\frac{F}{M^2 \lambda}Q + \mathcal{O}(Q^2).
\end{equation}
The final ingredient is the exact form of the dependence of $Q$ on $\Phi$. 
With the effective brane charge given by
\begin{equation}
Q(\Phi) = \pm \frac{(\Phi^N-\Phi_*^N)^K}{M^{N K -2}},
\label{eq:Q}
\end{equation}
where $N$ and $K$ are positive and integer valued parameters and the sign of the charge is not fixed, the difference in the Higgs VEV for neighboring vacua vanishes for $\Phi \rightarrow \Phi_*$.  Correspondingly, the density of vacua diverges at that point.
The result is a probability distribution for $\Phi$ with singular peak around $\Phi_*$. 
This point is called an attractor since given infinite time $\Phi$ will inevitably move arbitrarily close to $\Phi_*$. 

For derivation of equations (\ref{eq:F}-\ref{eq:Q}) and further details we refer to the original papers \cite{DV0304, D0410, D1908}.
In the following sections the terms \textit{brane nucleation} and \textit{timestep} are used interchangeably and $\Phi$ and $\lambda$ are measured in units of Planck mass $M_P$.

\section{Random Walk Model}
\label{sec:rwModel}
Since we are solely interested in the efficiency of the attractor mechanism we neglect all contribution to the Higgs mass other than from $F$, so we set $m=0$. With that the change of $\Phi$ simplifies to
\begin{equation}
\Delta \Phi = \frac{Q}{\lambda M^2} = \pm \frac{(\Phi^N-\Phi_*^N)^K}{\lambda M^{N K}}.
\label{eq:DeltaPhi}
\end{equation}
Sufficiently close to the attractor back-reaction on the inflationary background can be ignored, therefore the probability of nucleating a brane or an anti-brane can be assumed to be equal. Correspondingly we assign to moving either in positive or negative direction the probability $P= 0.5$.

In the following we will therefore model the time evolution of $\Phi$ as a symmetric random walk.
This is an extremely good approximation. Because brane nucleation is a non-perturbative process and rare, the nucleation of subsequent branes in the new vacuum is not sensitive to the original direction of nucleation. In other words, by the time that a given nucleation happens the walls of the bubble from the previous nucleation are exponentially far away and do not affect the succeeding nucleation direction.
 
We further simplify the original expressions slightly by assuming $\Phi_ * = 0$, merging the coupling constant $\lambda$ and the cutoff $M$ into one parameter $\mu$ and keeping the exponent of this parameter independent of $\nu \equiv NK$. The change $\Delta \Phi$ at each discrete timestep  (or equivalently brane nucleation) is then given in compact form by
\begin{equation}
\Delta \Phi(\tilde{\Phi}, \mu, \nu) =
 \begin{cases} 
       + \mu \tilde{\Phi}^\nu,& P = 0.5 \\
       - \mu \tilde{\Phi}^\nu,& P=0.5
   \end{cases}
   \label{randomStep}
\end{equation}
where $P$ denotes the probability of the specific outcome and $\tilde{\Phi}$ is the current value of the variable $\Phi$. For the remaining free parameters we assume $\mu \in (0,1)$ and $\nu$ is an integer and greater than one.

With that the random walk can be defined iteratively by
\begin{equation}
\Phi_{i+1} = \Phi_i + \Delta \Phi(\Phi_i, \mu, \nu),
\label{step}
\end{equation}
where $i$ indexes the brane nucleations and $\Phi_i$ denotes the value of $\Phi$ after the $i$th brane has nucleated. 

It is straightforward to generalize this model to multiple different charges on the brane. Since these are independent from each other, the sequence in \Eq (\ref{step}) can be generalized to  
\begin{equation}
\Phi_{i+1} = \Phi_i + \sum_{k=1}^d \Delta \Phi(\Phi_i, \mu_k,\nu_k),
\label{randomWalk}
\end{equation}
for $d$ distinct charges and possibly different and independent parameters $\nu_k$ and $\lambda_k$.
So at every timestep the VEV of the Higgs is shifted by $d$ different terms where again the sign of each contribution is equally likely due to the unfixed signs of the individual charges.

\section{Simulations}
\label{sec:main}
The random walk defined by \Eq \eqref{randomWalk} can be simulated on a computer. Of course, due to the statistical nature of the process, quantitative analyses can only be performed with a high number of realizations. The main focus here should lie on the question how the number of distinct brane charges $d$ affects the convergence rate of the random walk. 

\begin{figure}[ht]
	\centering 
	\begin{subfigure}{0.49\textwidth}
	\includegraphics[width=\textwidth]{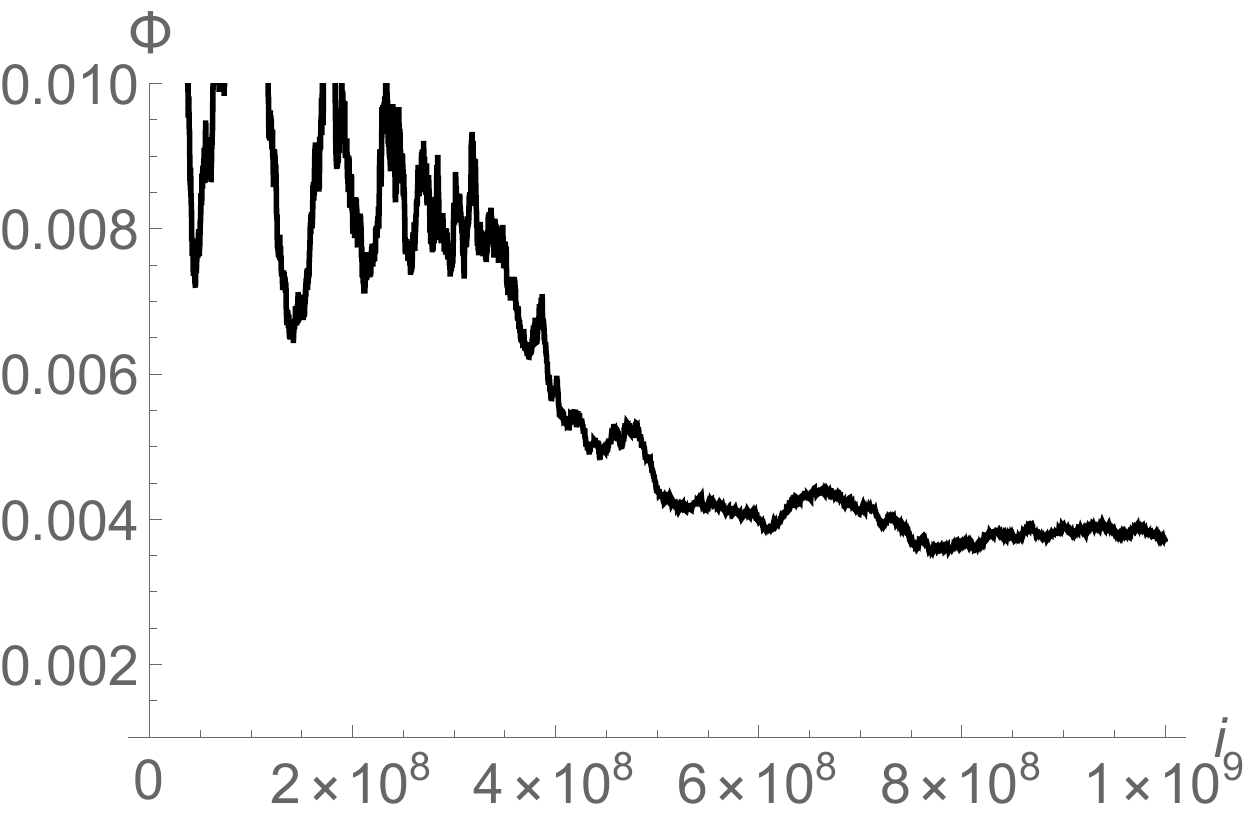}
	\caption{$d=1$}
	\end{subfigure}
	\begin{subfigure}{0.49\textwidth}
	\includegraphics[width=\textwidth]{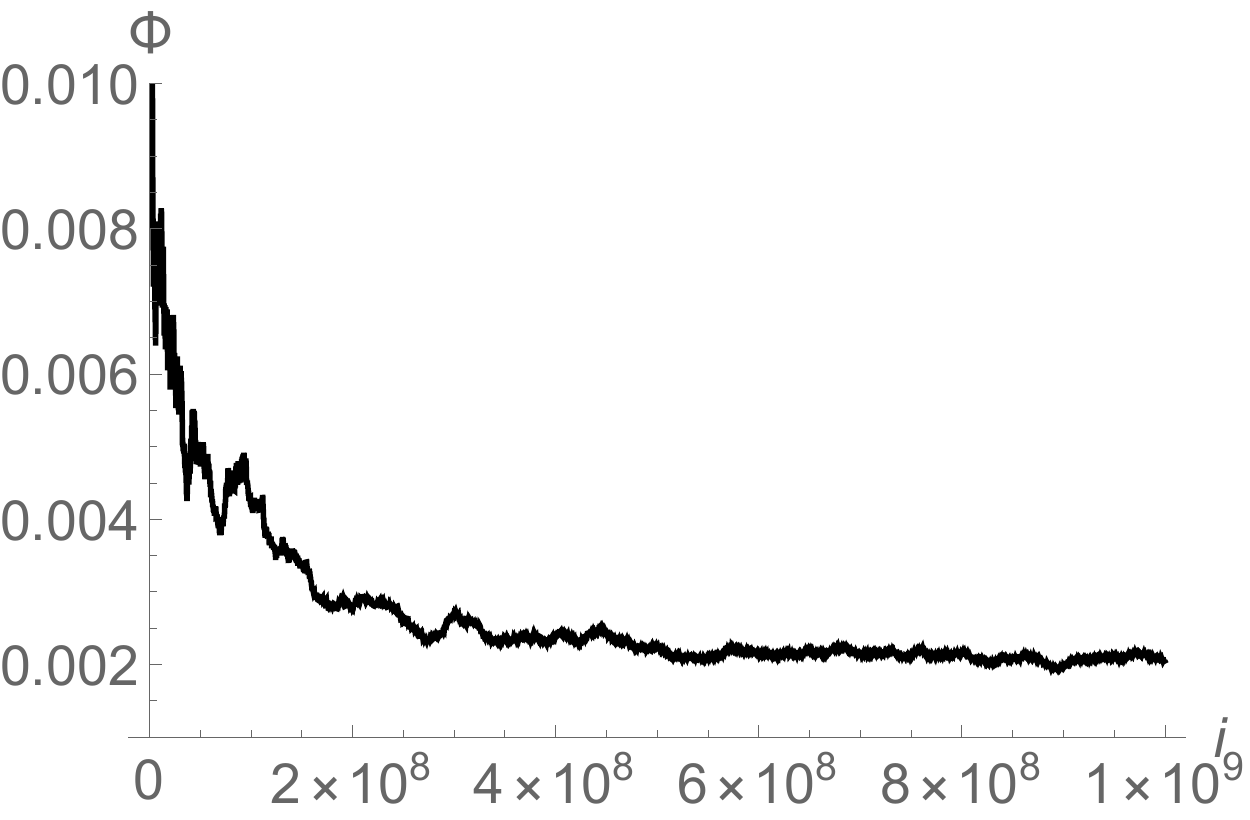}
	\caption{$d=10$}
	\end{subfigure}
	\caption{Exemplary realizations of the random walk defined in \Eq \eqref{randomWalk} with $\mu = 0.5$, $\nu =3$ and $\Phi_0 = 0.1$  for different number of brane charges. The plot range has been set to emphasis late time behavior. In this specific example the run with higher number of charges converges faster to smaller values compared to the $d=1$ case.}
	\label{realplots}
\end{figure}

All simulations in this section have been performed with an initial value of the Higgs VEV given by $\phi_0 = 0.1$ and the two free parameters were set to $\lambda = 0.5$ and $\nu =3$, respectively and equal for all three-form fields. Every simulation was calculated for $n = 10^9$ steps. For late times we plot illustrative random walks for $d=1$ and $d=10$ in \fig \ref{realplots}. Matching the intuitive picture, both walks decrease on average as a function of the timestep $i$. Of course, the smaller $\Phi$ becomes, the slower becomes the rate in accordance with \Eq (\ref{randomStep}). Thus for the Higgs VEV to relax infinitely close to zero, infinite time would be required. However, this time requirement can be accounted for by eternal inflation \cite{Vilenkin:1983xq, Linde:1986fd} as already explained in section \ref{sec:review}.

\begin{figure}[!ht]
	\centering 
	\includegraphics[width=0.65\textwidth]{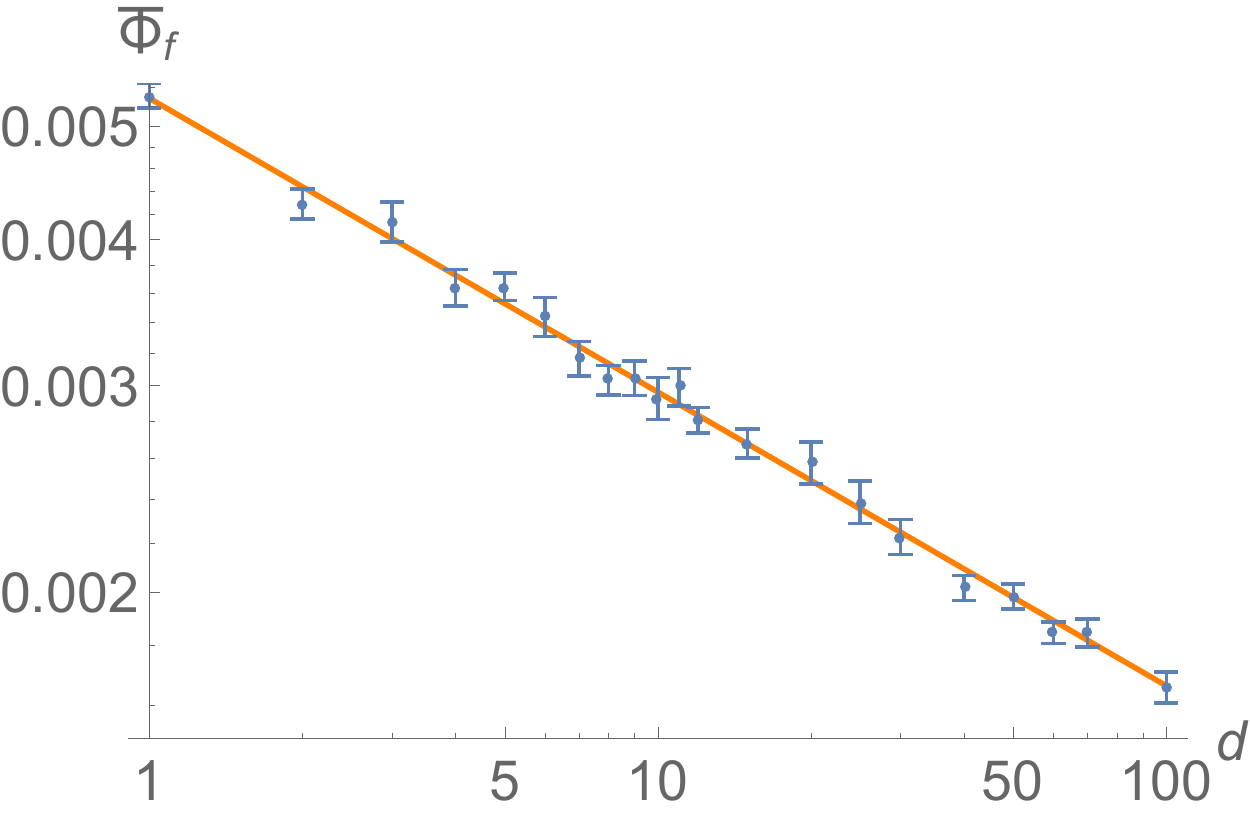}
	\caption{Average final value $\overline{\Phi}_f$  after $n=10^9$ timesteps as a function of the number of charges $d$. For every point the average has been taken over 100 runs. For a fit function of the form $a \cdot x^b$ the parameters attain following values: $a=0.0053$ and $b=-0.25$ with standard errors $\sigma_a = 5.3\cdot 10^{-5}$ and $\sigma_b = 0.005$. The parameters of model (\ref{randomWalk}) have been set to  $\mu = 0.5$ and $ \nu = 3$ with an initial value given by $\Phi_0 = 0.1$. The standard error is depicted as error bars around its mean value.}
	\label{convRate}
\end{figure}

Next, we analyze the dependence of the convergence rate of $\Phi$ on the number of charges $d$. For this purpose we average over the final value $\Phi_f$ after $n=10^9$ steps for 100 runs\footnote{Runs that diverged at some point ($\Phi_i > 1$ for some $i$) were discarded and reset.} for various $d\in [1,100]$. The corresponding data is plotted in \fig \ref{convRate} in blue. In accordance with \fig \ref{realplots}, we clearly observe an enhanced relaxation efficiency for higher $d$. 

For quantitative conclusions we fit the data in \fig \ref{convRate} with a function of the form $a \cdot x^b$ in orange. The fit parameters and their respective standard deviation are presented in table \ref{table}. The exponent matches, within statistical errors, the averaged exponent when fitting the values of $\Phi$ for individual runs. This completely matches the analytic intuition that increasing the brane nucleation channels is equivalent (up to a numerical factor) to the nucleation of more branes. The only difference that can be observed is at the beginning when $\Phi$ is still large and when the attractor's pull is the strongest. Greater $d$ results in larger jumps within $n'$ steps in comparison with an equivalent run with $n''=d \cdot n'$ steps and only one charge. For later times and small $\Phi$, however, the difference between jump distances for neighboring values of $\Phi$ becomes negligible. 

We repeated the analysis also for $\nu \in \{2,4\}$\footnote{Within the considered number of steps the numerical value of $\Delta \Phi$ with  $\nu \ge 5$ can not be resolved accurately enough with standard machine precision. However, this problem can of course in principle be solved and we don't expect any significant deviation from our results in this regime.} The results are presented in table \ref{table}. We observe again that the inclusion of more brane charges is up to a numerical factor equivalent to additional nucleation of branes. We also checked for $\nu=3$ that varying $\mu \in [0.05,0.5]$ only results in a change of the prefactor $a$. We expect this behavior to hold also for smaller values of $\mu$.
\begin{table*}
\centering\ra{1.3}
\begin{tabular}{l|cccccccc}\toprule
$\nu$ & $a$ & $\sigma_a$ & $b$ & $\sigma_b$ & $\tilde{a}$ & $\sigma_{\tilde{a}}$&$\tilde{b}$ &$\sigma_{\tilde{b}}$ \\ \midrule
2 & $4.8 \cdot 10^{-5}$ & $1.1 \cdot 10^{-6}$& -0.48 & 0.014&0.05& 0.02&-0.5& 0.04\\
3 & 0.0053&$5.3 \cdot 10^{-5}$&-0.25&0.005&0.17&0.07&-0.25&0.06\\
4 &0.026 & $2 \cdot 10^{-4}$ & -0.17 & $2.7 \cdot 10^{-3}$ & 0.25 & 0.1 & -0.16& 0.04
\\\bottomrule
\end{tabular}
\caption{Fitting values for the evolution of $\Phi$ as a function of brane charges and number of nucleated branes. The values for the ladder are marked by a tilde. For both cases the fitting function is of the form $a\cdot x^b$. The standard errors are denoted by $\sigma$.}
\label{table}
\end{table*}

\section{Conclusions}
\label{sec:end}
In this paper we have studied numerically the model first introduced in \cite{DV0304, D0410} which solves the Hierarchy Problem by cosmological relaxation of the Higgs mass towards the attractor vacuum during eternal inflation. 
In this scenario the Higgs mass/VEV is changed due to nucleation of branes (or axionic domain walls). At the same time the Higgs VEV acts as a back-control parameter that directs the convergence of the relaxation progress. 

Such an attractor could in principle be realized in various different models each with its own specifics. In this analysis, however, we only focused on universal key features of the Cosmic Attractor mechanism. For this we have modeled the Higgs VEV evolution as a random walk with each step mimicking a vacuum transition triggered by a brane nucleation.
The observed convergence to $\Phi_* = 0$, which represents our attractor point, is in accordance with analytic considerations. 
We then generalized this stochastic model to multiple different charges sourced by a 2-brane and studied the impact of their number on the relaxation rate of the Higgs VEV.
We showed that a higher number of three-form fields leads to a faster convergence rate. That is less brane nucleation are necessary to relax the Higgs vacuum expectation value below a given value. This confirms the intuitive picture that adding brane charges is equivalent to an increase of brane nucleation channels.

\section*{Acknowledgments}
We thank Gia Dvali for valuable discussions throughout the project and helpful comments on the manuscript. We are also very grateful to Georgios Karananas for insightful discussions.

\end{document}